\documentclass{revtex4}
\usepackage{amsmath}
\usepackage{graphicx}
\usepackage{booktabs}
\usepackage{supertabular}
\usepackage{color}
\usepackage{hyperref}
\usepackage{caption}
\usepackage{float}

\begin{document}
\author{Behnam Parsaeifard}
\affiliation{Physics department, Sharif University of Technology, P.O. Box 11155-9161, Tehran, Iran}

\author{Saman Moghimi-Araghi}
\affiliation{Physics department, Sharif University of Technology, P.O. Box 11155-9161, Tehran, Iran}

\title{Controlling Cost in Sandpile Models Through Local Adjustment of Drive}

\begin{abstract}
        In this paper we consider sandpile models and modify the drive mechanisms to control the size of avalanches. The modification to the drive mechanism is local. We have studied the scaling behavior of the BTW and Manna models. We have found that the BTW model is more sensitive to the modification than the Manna model. Furthermore we have assigned a cost function to each avalanche and have found an optimum value for the modification to arrive at the lowest cost.
\end{abstract}
\maketitle
\section{Introduction}

     The concept of self-organized criticality (SOC) proposed by Bak, Tang and Wiesenfeld \cite{BTW}, has attracted a lot of attention as a possible general framework for explaining the occurrence of power laws in nature. Power laws are usually associated with criticality in usual thermodynamical systems, where it needs fine tuning of external parameters to arrive at criticality. Yet there exist many phenomena in nature which show power law behavior without tuning any external parameter. These phenomena are supposed to fall in the set of SOC systems. Earthquakes \cite{EQ}, forest fires \cite{FF}, water reservoirs \cite{WR}, human brain \cite{HB} and financial markets \cite{FM} are proposed to be a few examples of such phenomena. In these systems, although the interactions are short-ranged; the correlation length is of the order of system size and the chance of system-wide catastrophe is finite; for example in nature very strong earthquakes are not too rare.  

In typical SOC systems, it is usual to observe cascading failures. In such a cascading event,  a random failure in a small area of the system leads to a chain of ongoing reactions that can eventually cause collapse of a large fraction of the whole system. The cascading failures are observed in many systems, from power transmission systems \cite{PG} to economical \cite{econ} and biological systems \cite{biol}.    
Thus it is very interesting and challenging to control SOC systems and as these systems are complex and usually have feedback mechanisms it is usually difficult to predict the respond of the system to perturbations. The main goal of controlling  such systems is to  decrease the systemic risk; that is to prevent system-wide catastrophes or very large events as much as possible. For example in power grids it has been shown that suppressing small blackouts may lead to a higher risk of large ones \cite{PG}. 

In this paper, we will consider two well-known sandpile models as our SOC system: namely the BTW and Manna models. We introduce a local control to the system and investigate how the two models are affected by this perturbation.  We first examine how critical behavior of the systems changes and then, by defining a cost function both for the events of SOC system and level of our perturbation, we will find an optimum value for the perturbation level to minimize the cost. The paper is organized as follows: we will give a very short review of the BTW and Manna models in section II. In section III we will introduce our perturbing mechanism and investigate the scaling behavior of the models under such perturbation. In section IV we will compute the cost functions of the systems and find the optimum drive for them.  

\section{Introduction to BTW and Manna Models}\label{ASM}
The BTW model is defined on a two dimensional square $L\times{L}$ lattice. To each site on the lattice, a height variable, $h_{i}$, is assigned. These variables  can take any value from the set $\lbrace{1, 2, 3, 4}\rbrace$ and usually are regarded as the number of grains of sand on each site.  In each step, a random site  $i$ is selected and a grain of sand is added to it; that is, $h_{i}\rightarrow{h_{i}+1}$.  If the resulting height is equal or less than 4, another site is chosen and a grain of sand is added to it. However if the height of a site becomes more than 4, the site becomes unstable and topples: it loses 4 grains of sand, each of which is transferred to one of the four neighboring sites. This may cause the neighboring sites become unstable and topple and as a result a chain of topplings may occur through the system, which is called an avalanche. During the toppling process, the number of sand grains is conserved, however if a boundary site topples, one or two grains of sand leave the system. This process continues until the system reaches  a stable configuration in which there exist no unstable sites and the avalanche comes to an end. At the next step a grain of sand is added to another randomly selected site and the system evolves as explained before. After a finite number of steps, the movement on the space of stable configurations leads the system to a subset of configurations  which is called the subset of  recurrent states. 

Every avalanche is characterized by a set of parameters. The size of an avalanche is the number of topplings occurred within the avalanche while the area of avalanche is the number of distinct sites that topple. The duration of an avalanche is equal to the
number of update sweeps needed until all sites are stable again. It has been found numerically that in the steady state the corresponding probability distributions obey power-law behavior and the system shows finite-size scaling, therefore it is critical\cite{IvashkevichKtitarevPriezzhev,DharManna,LubeckUsadel,KitLubGrasPri,DemenechStellaTebeldi}. 

The Manna model is defined on the same lattice \cite{Manna}. In this model the height variables can be either zero or one. At each step a grain of sand is added to a randomly selected site. If the height variable becomes more than one, it topples in the following way: it loses two grains of sand and these two grains are transfered to two randomly chosen neighbors, in a specific model which we will consider in this paper, the two grains of sand are transfered either to the left and right neighbors or to the up and down neighbors with equal probability. Therefore the process in which the avalanches are made is not deterministic anymore.

The same parameters for avalanches can be defined in the Manna model. Again the probability distribution of the corresponding parameters show power-law behavior and the system has finite size scaling. The exponents in the Manna model and BTW model are somehow different.  In fact, the precise identification of universality classes in sandpile models has been a controversial issue\cite{BenHur,Chessa,AMN}. In the original paper \cite{Manna} it was indicated that stochastic models are in the same universality class as the deterministic models. This suggestion was supported by some heuristic and analytical arguments \cite{VespZap95,Nakanishi97,LubeckHeger,Pietro94,ChessaVespZap99}. However,  subsequent studies that focus on an extended range of observables suggest that the two model belong to different universality classes \cite{Lubeck2000,DickmanCampelo,StellaDemenesh01,Karmakar05}.

The other difference of the two models is the presence of multi-scaling in BTW while the Manna model shows a clear single-scaling behavior. Tbaldi et. al. \cite{Tbaldi99},  doing a moment and spectra analysis, showed that in BTW the scaling relations can not be expressed by a single exponent, in other words it shows multi scaling. Later De Menech and Stella \cite{DemenechStella} showed that the so called waves of toppling have  simple scaling properties and in the Manna model the statistics of waves is the same as avalanches, while in BTW it is not true. We will see that the multi-scaling property of BTW model is amplified as we change the drive method in a certain way.

In this paper our main concern is not the strict scaling properties of the model. Our main goal is to give an estimate of cost as a result of avalanches and find an optimum external drive method. Therefore in the next section we just qualitatively explore the critical behavior of the models under the introduced  local  change of the drive and investigate if the critical features of the model fade away or not.

\section{Modification of drive mechanism and its effects on critical behavior of the sandpile models}

As the Self-Organized Critical systems have fat-tail statistics, large and catastrophic events are not too rare. Avoiding such large events would have significant benefits, therefore it is an interesting question whether it is possible to change the critical features of the system through some modifications. The modification can be made in different ways \cite{noel,Caj,Agos,Winkler,Pfen,Bhau}, for example one can  modify the dynamics or the dissipation mechanism \cite{Caj,Agos}.  In this paper we do not change the dynamics of the system, rather we  consider the modification of the method the system is driven. Such an approach has been considered before \cite{Winkler,noel}. In \cite{Winkler} BTW model is driven in the following way: The random site to which the sand grains are added is only selected among the sites that have a specific height $n$. They have considered different $n$'s and investigated the critical behavior of the system and have found out that the usual critical properties of the system vanishes except when the sites with maximum height are favored. 

In \cite{noel} the drive is changed in another way. They have considered a sandpile model on a random 3-regular graph in which all of the sites had three neighbors and their heights could take the values 0,1 and 2. Their  system doesn't have boundary; Therefore a bulk dissipation controlled with the parameter $\epsilon$ is introduced to the model. They drive the system using the following method:  the probability that a new grain of sand is added to a site with the maximum height $h=2$, is controlled with a parameter $\mu$. Note that the sites with $h=2$ are those sites that will become unstable if a grain of sand is added to them.

A cost function was defined in the system which depended on the size of the avalanches occurred. Then they investigated how the cost changes as $\mu$ is increased and through this procedure the optimum value for $\mu$ is obtained that minimizes the total cost. In their research they have assumed that if adding a grain of sand does not cause an avalanche, not only there is no cost, but also there will be a certain amount of benefit. In this way, for some values of $\mu$ the system had an overall benefit (negative cost) which does not seem to be an expected result. In other words occurring a very small avalanche is not so different from having an avalanche with zero size. One may think of earthquakes where, the occurrence  of a small earthquake has no cost or benefits comparing to the case in which no earthquake has happened at all. Additionally such a change in driving the system needs some non-local information of system, like the total number of sites with height 2 which may not be at hand for a local observer.

In our model, we modify the drive method in a way that only local information of the system is needed: in each time step a site is selected randomly. Before adding a grain of sand to the site, we check the height of the site and its 8 neighboring sites. Then the sand is added to the site among these 9 sites that has the minimum (or in an alternate modification has the maximum) height. If more than one sites share the minimum (maximum) value of height, we randomly select one of them and add the grain of sand to it. We call this kind of driving the system minimum-drive or maximum drive depending whether we have picked the minimum height or the maximum height.  In the following we apply these rules to both BTW and Manna models and investigate how the critical behavior is changed. For simplicity we call BTW model driven by minimum/maximum-drive, Min model/Max model and the Manna model driven by minimum/maximum-drive, Manna Min model/Manna Max model.
\vspace{5mm}

\begin{figure}[h]
\centerline{\includegraphics[scale=.33]{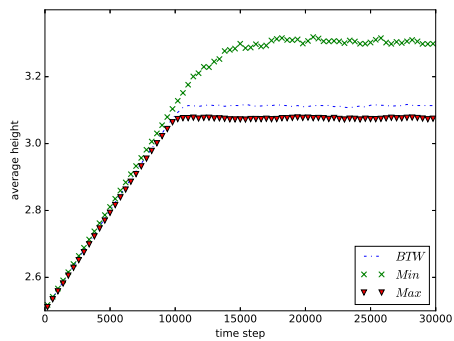}}
\caption{The average height of ordinary, Min and Max BTW model in a $512\times 512$ square lattice. At the beginning all the sites have random heights. }
\label{averageBTW}
\end{figure}

\subsection{ BTW Model}

Consider the BTW model on a $L\times L$ square lattice. It is naturally expected that in minimum-drive the number of the sites with smaller heights be less than the ordinary BTW model. This is because when we find a site with a small height in the neighboring of a selected site, we will add sand grains to it. In contrary, in maximum-drive the sites with larger heights are expected to be fewer than that in ordinary BTW model with the same reasoning. Fig. \ref{averageBTW} shows the average heights of the three models as functions of time: the ordinary BTW model and Min and Max models. Initially, all the sites were considered to have a random height.

It is observed that in the case of Min model, the difference of the average height with ordinary BTW model  is more clear, and the plateau part (which corresponds to the steady state of the model) fluctuates more than the ordinary case. However the change of average height is not very significant in Max model. As an important parameter, one can mention the probability of finding a site with height 4 which is about 0.49 in minimum model, about ten percent more than ordinary BTW system (0.44), but still less than the percolation threshold in square lattice ($\simeq 0.59$). If this probability reaches the percolation threshold, most of avalanches would be of the order of the whole system. In maximum-drive case this probability is about 0.42 which is a bit less than the value in ordinary BTW.

\begin{figure}
\centerline{\includegraphics[scale=.40]{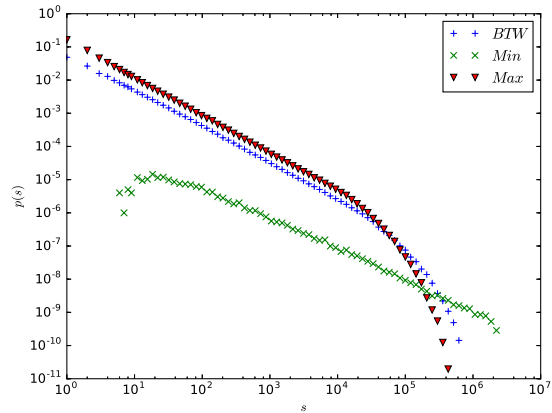}}
\caption{The probability distribution function of ordinary, Min and Max BTW models in a $256\times 256$ square lattice. }
\label{pdfBTW}
\end{figure}

We will now turn to the probability distribution of the avalanche sizes to investigate the scaling behavior of the system to see if they are still at the critical state or not,  and if they are, is the universality changed. The changes in probability distribution of the avalanche sizes is naturally expected to happen. For example in Max model, we always add sand grains on sites with higher heights, and as said before,  effectively we reduce the number of the sites which have maximum height and are ready to topple with minimum stimulation. Therefore we expect that the number of large avalanches is reduced. On contrary, in Min model, we try to avoid making avalanches and add sand grains to the sites that will not be activated by this addition. Therefore, gradually the number of sites having maximum height is increased and if an avalanche happens, it will have a good chance to spread out in the system. Figure \ref{pdfBTW} shows the probability distributions of the avalanche sizes of the three systems. The system size is 256 and after reaching the steady state, 10 million sand grains are added. It is observed that although the curve associated with the Max model, has been changed in a way that smaller avalanches are more probable, but it does not have significant difference from BTW model's curve. The case of Min model is completely different. First of all, the probability distribution function (PDF) has a maximum at $s\simeq 14-16$. Additionally for most of avalanche sizes, the probability distribution function is significantly smaller than ordinary BTW's case; at some points more than $10^3$ times smaller. Instead the cutoff of the probability distribution is extended from $10^4$ in ordinary BTW to about $10^6$. Note that the total number of sites in this lattice is about 65000 and having an avalanche of size $10^6$ means that every site has toppled more than ten times on average. 

Before discussing scaling properties of these models, we are going to find the point at which the PDF of Min model shows a maximum. If we look more closely at figure \ref{pdfBTW} we can see that basically there exist no avalanches  with size less than 4. In fact when we choose a site to add a grain of sand to it (or to its neighbors), an avalanche occurs  only if all its neighboring sites have 4 grains of sand and are ready to become unstable: Even if only the height of one of them is less than four, that site would be selected and a grain of sand would be added to it, and hence no avalanche would occur. Therefore an avalanche occurs If all 8 neighbors (as well as the central site)are at the maximum height, and in this case, the size of the resulting avalanche is at least 10  (the central site topples at least twice). Because of boundary sites, avalanches of size less than 10 may occur.  As the number of boundary sites is small, these avalanches are rare and become negligible in large systems.

\begin{figure}
\centerline{\includegraphics[scale=.70]{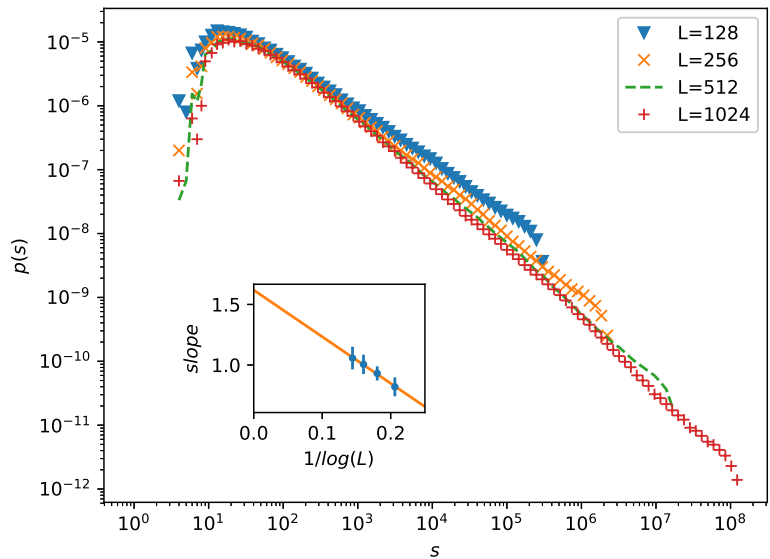}}
\caption{Probability distribution function of avalanches of Min model for different system sizes. No strict power law with a single exponent is found in these PDF's.  }
\label{min-L}
\end{figure}   

The more important question is whether the new models  are still critical or not. Let us consider the  Min model first. Figure \ref{min-L} shows the probability distribution function of Min model with different system sizes. As expected, as the size of the system ($L$) is increased, the small-sized avalanches (avalanches with size less than 10) are much less likely to happen. After the peak, there exists a more or less linear part on the PDF, however  the slope of this linear part increases substantially as we move on toward the large avalanches in the graph. For example in the case of $L=512$ it varies between  $0.97$ for avalanches with size less than $10^4$ and $1.14$ for avalanches with size grater than $10^4$. The same phenomena is observed in all system sizes, additionally the exponents vary from one system size to another. As an example for the system with size $L=256$ the slope varies between $0.86$ to $0.96$. Therefore, a single scaling behavior is not present at the model. In fact the multi-scaling property of the BTW model is amplified with our drive method. Let us for the moment consider an average value for the slope and call it $\tau_L$. As in the ordinary BTW model, we  observe that $\tau_L=\tau_\infty-c/\log L$ (inset of fig.\ref{min-L}) with $\tau_\infty=1.6\pm 0.3$. The large error-bars of individual points leads to even larger error-bar in $\tau_\infty$ as we extrapolate $L\rightarrow\infty$.  Note that the value of $\tau_\infty$ of ordinary BTW ($\simeq 1.25$) is consistent with min model within the error-bars, although the PDF of avalanche sizes clearly shows multi-fractal property. We have also done a finite-size-scaling analysis of the system: although there is no clear data collapse, yet the avalanche size that the PDF deviates from a linear curves obeys a scaling law $s_c\sim L^D$ with $D=2.8\pm 0.1$, consistent with the exponent of original BTW model.

Now we turn to Max model: The PDF of Max model doesn't differ from the BTW model significantly. Although the PDF of Max model does not coincide the usual BTW model, the model shows finite size scaling and the exponents are just the same as the ordinary BTW model. Using computer simulations for a variety of system sizes, we have done finite size scaling (FSS) analysis on the system and have obtained critical exponents to be $\tau=1.28\pm 0.01$ and $D=2.7\pm 0.1$ where $\tau$ and $D$ are defined through the scaling relation $p(s)=s^{-\tau} f\left(s/L^D \right)$.

\subsection{ Manna model}

In this section, we study the effects of new drives introduced in the previous section, on the Manna model. Similar to the BTW model, we define Min/Max Manna model as following: in the Min/Max Manna model, a site is selected randomly, and a grain of sand is added to a neighboring site with smallest/largest number of sand grains. If there exist more than one neighbor having the same minimum/maximum height, sand grain is added to one of them randomly. 

\begin{figure}
\centerline{\includegraphics[scale=.45]{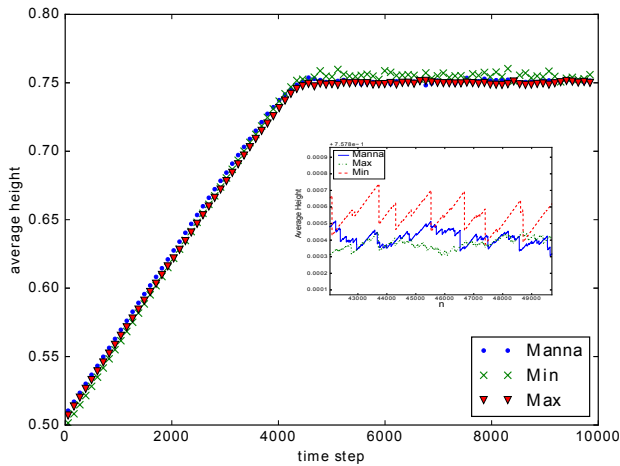}}
\caption{Average height of the Min Manna and Max Manna model for L=256. The mean-height of all models is the same. }
\label{mannameanh}
\end{figure}

As in the case of the BTW model, it is expected that in steady state of the Min Manna model the number of empty sites (the sites with no grains) be fewer than the number of occupied sites (the sites with height equal to 1) and hence  the average height in the Min model be more than the ordinary Manna Model. Also with a similar reasoning it is expected that the average height of the Max Manna model be less than the ordinary model. However, it is observed that average heights of the three models are nearly the same (see figure  \ref{mannameanh}, the main graph). Although they are very similar, if we zoom in, some differences are observed.  In all the models the average height grows gradually but it abruptly falls down resulting a sawtooth-like shape as seen in the inset graph of figure \ref{mannameanh}. The amount of raise is different in the three models: the Min model has the greatest raises and the Max model has the smallest one.
 
\begin{figure}
\centerline{\includegraphics[scale=.5]{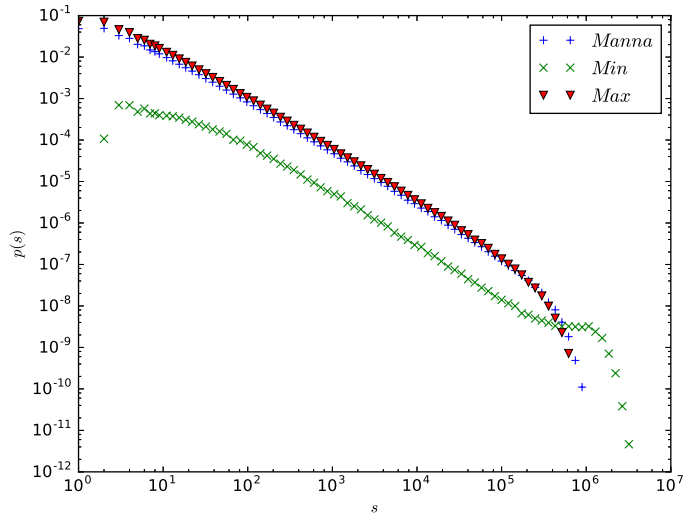}}
\caption{Comparison of probability distribution function of the Min, Max, and Manna models for $L=256$. The PDF of Min model extends to much larger avalanches.}
\label{mannapdf}
\end{figure}

Let's now examine the critical behavior of the models. We have simulated Min and Max models for different system sizes. Figure \ref{mannapdf}
 shows the distribution function of avalanches for the Max, Min and ordinary Manna models for $L=256$. In spite of the fact that the average height for these models is the same, it is very interesting that the PDF of Min model is completely different. Just as in the case of the BTW model, Max Manna model is very similar to the ordinary Manna model and deviates only slightly from it.  

In the Min Manna model, the smaller avalanches are less probable while the cutoff of the scaling regime has been extended to larger values. Avalanches with very few topplings are rare, this is due to the drive method: an avalanche occurs only if we have a $3\times 3$ square in which all the sites are occupied. 

\begin{figure}
\centerline{\includegraphics[scale=.5]{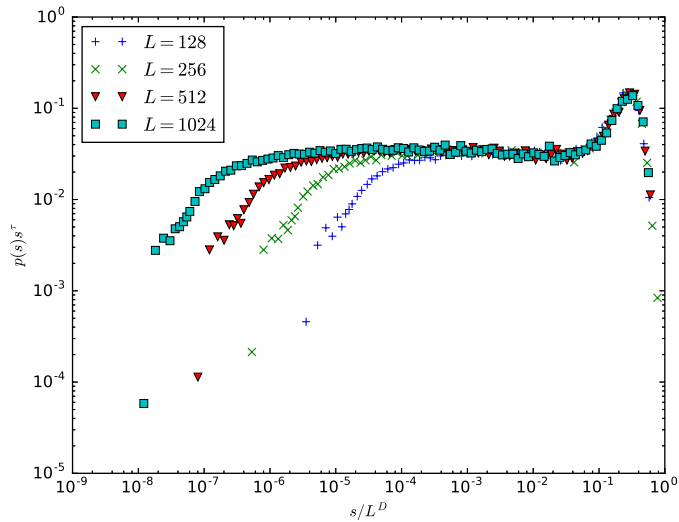}}
\caption{The Min Manna data collapse. The parameters $\tau$ and $D$ are set to $1.27$ and $2.73$ respectively.}
\label{minmanna}
\end{figure}

In the BTW model, Min drive caused the system to reveal its multi-fractal more explicitly.  However, in the case of the Manna model it seems that Minimum-drive does not ruin the critical behavior at all. Using FSS analysis, we investigated the PDF of the Min Manna model. Figure \ref{minmanna} shows the data collapse for different system sizes, from which It is clearly observed that the system is still critical. Obtained values for $\tau$ and $D$, are $1.27\pm 0.01$ and $2.73\pm 0.05$ respectively, which are close to those obtained for the ordinary Manna model. 

An interesting feature of the Manna model is its robustness to perturbations. It has been also seen that average height was not shifted significantly by changing the drive. In Min model, the way we drive the system tries to fill the empty sites; therefore it is very probable that after a while we find a region which is completely occupied. Let us see what happens if we add a grain of sand to a region in which all the sites have height 1. We take the region a square of size $(2m+1)\times(2m+1)$ and add grains of sand to the site at the center. For simplicity we suppose that the boundary sites are open and the grains of sand do not return to the region after they leave it. To compare, we consider the same experiment with the BTW model, that is we consider a square of size $(2m+1)\times(2m+1)$ of sites with height equal to 4 and add grains of sand to the centering site.  In both BTW and Manna cases, it is expected that the mean height gradually fall down to its value at steady state (3.125 and 0.75 respectively). In the case of the BTW model the problem can be tracked analytically. For example after adding the first grain of sand and letting the system relax, one arrives at the configuration where all the sites except the sites on the diagonals of the square have height equal to 4. In other words, if $m$ is relatively large we will have: $p(4)\simeq 1-O(1/m)$ and the change in the average height is small. However, in the Manna model we  observed numerically that just after adding one grain of sand the average height of the system arrives nearly at its steady state value (0.75). Thus, it seems that the dynamics of the Manna model is very robust to perturbations that try to increase its average height while the BTW model is not. Also the sawtooth shape observed in figure \ref{mannameanh} can be explained in this way. In conclusion, changing the drive in the way we did, the critical properties of the Manna are not affected.

\section{Optimization of cost function with respect to drive}

The main purpose of this article is to see how changes in the drive method could reduce catastrophic events. To this end, we need a parameter to tune the modification introduced to the system  and then by defining a suitable cost function we will be able to find an optimum value for the parameter to minimize the total cost. 

In our drive method, we compare the height of selected site with the height of its eight neighbors. We propose two distinct methods to tune the drive modification: (I) As the search is costly, we do it not every time, but with probability $\mu$. In this way we change the frequency of implementation of the modified drive. (II) We change the depth of search for sites with larger (smaller) height; that is, the search is done among $n$  neighbors of the original sites. There is no need to take $1\leq n\leq 8$, we may consider larger values for $n$ and search among the next nearest neighbors and further neighbors too. In the previous section, it was observed that Max drive could reduce the chance of occurring catastrophic events. Therefore, in this section we mainly focus on the models with Max drive. 

To give a measure of how much catastrophic an event is, we have to introduce a cost function for the avalanches. The cost clearly should be an increasing function of avalanche size, additionally the cost function is usually a concave function. This comes from the idea that a change in the size of  small avalanches does not lead to a large change in the cost. Similar to \cite{noel}, we take into account a cost function of the form 

\begin{equation}\label{costf}
A(s)=c s^\alpha,
\end{equation}
where $A$ is the cost, $s$ is the size of avalanche and $c$ and $\alpha$ are two constants. The parameter $c$ defines a scale for the cost and the parameter $\alpha$ should be taken to be more than one so that the cost function becomes concave. Such cost functions can also arise from risk aversion \cite{Newman}. The only difference between the cost function considered here and that of introduced in \cite{noel} is that no benefit is associated to avalanches of zero sizes in this paper which seems to be more reasonable.

To reduce boundary effects, It is possible to consider systems with periodic boundary conditions while imposing a small bulk dissipation $\epsilon$  to the system, that is, whenever the number of grains on a site exceeds the threshold and the site becomes unstable, it loses one grain of sand by probability $\epsilon$ and if it becomes stable then it does not topple. We have performed simulations with both methods. In the case of bulk dissipation  $\epsilon$ is set to the range $10^{-6} - 10^{-5}$ in order that the systems with sizes up to $L=1000$ be critical. With such a choice, the results turned out to be qualitatively the same.  The probability distribution function of avalanches, $p_{\mu,n}(s)$ is obtained numerically for different values of $\mu$ or $n$ and the average cost is calculated as a function of $\mu$ or $n$ via 
\begin{equation}\label{average cost}
\langle {\rm cost}\rangle =\int p_{\mu,n} (s)A(s)ds.
\end{equation}

\begin{figure}

\centering
	\includegraphics[scale=.40]{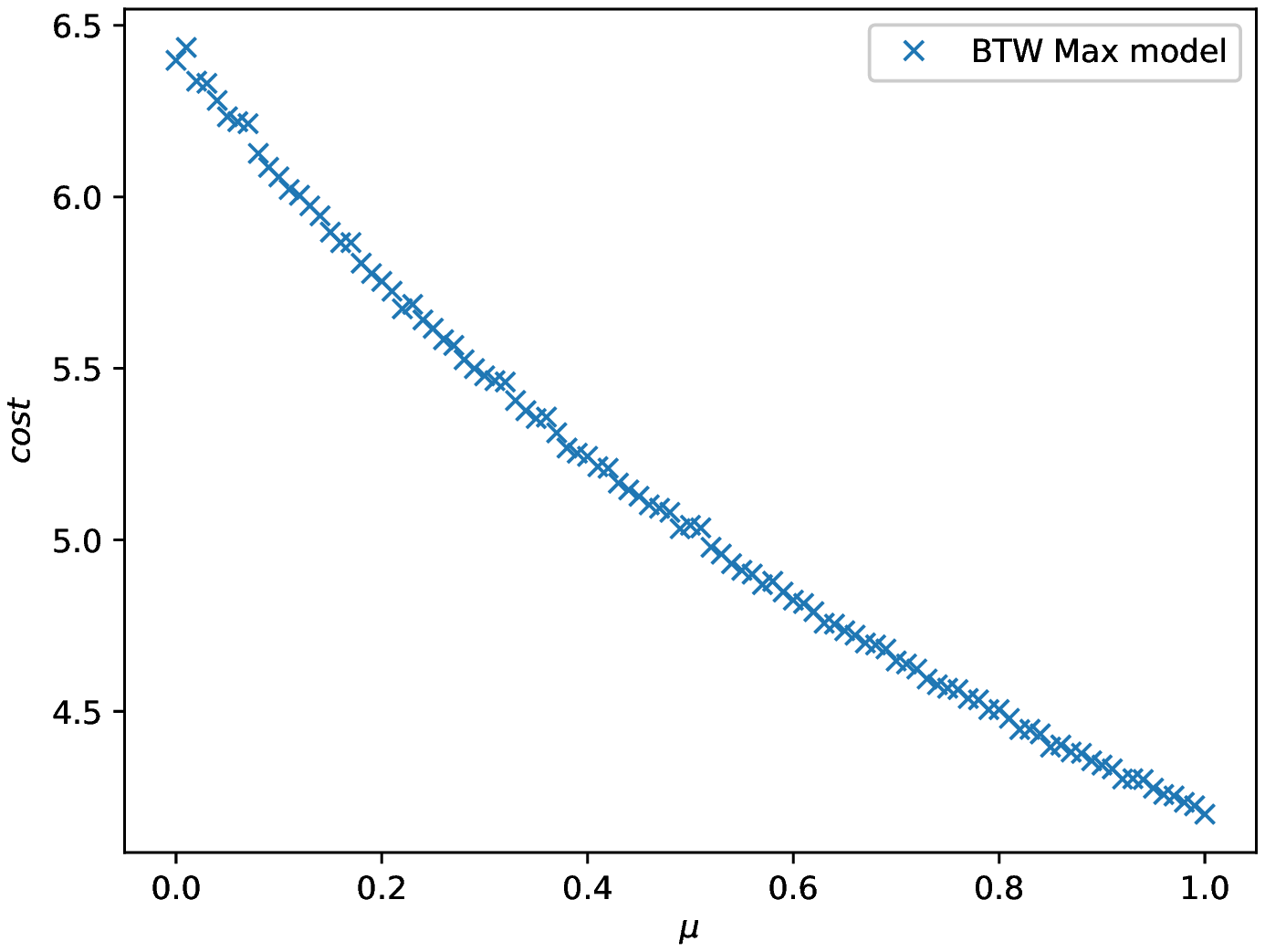}
	\includegraphics[scale=.40]{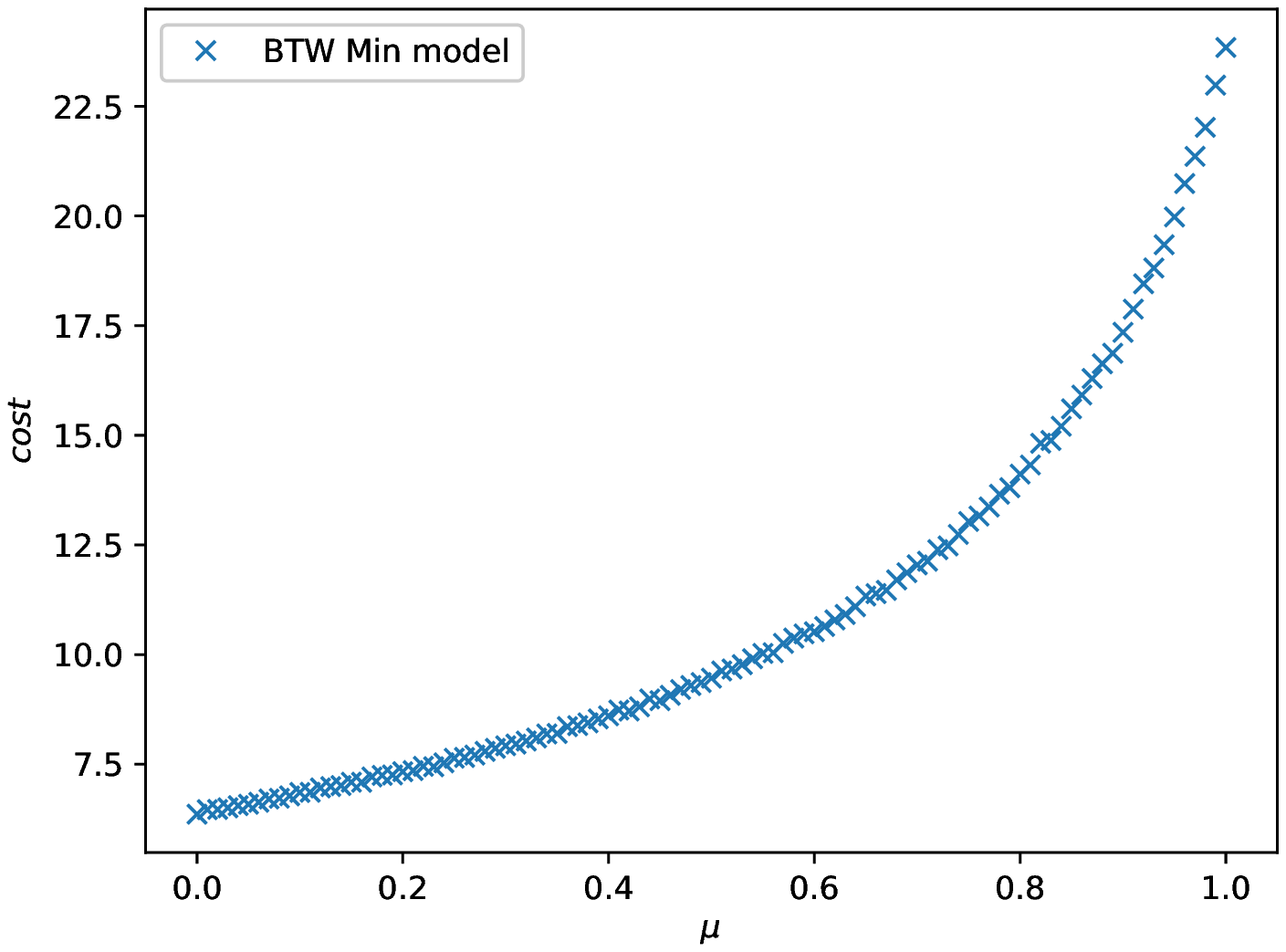}

\caption{Mean cost of the BTW system as function of $\mu$, the probability that Max and Min drive is implemented  In both cases $L=512$ and no cost is assigned to searching neighbors.}
\label{BTWcost}
\end{figure}

Let us first consider the case where the modification of drive method is applied with probability $\mu$. Figure \ref{BTWcost} shows the total cost as a function of $\mu$ for Min and Max BTW models respectively. Here we have considered $\alpha=1.5$ and $L=512$.  The vertical axis is in arbitrary unit which is set by the parameter $c$ introduced in the definition of cost function (equation (\ref{costf})). As expected, the cost for both model are the same when $\mu=0$. Also as the parameter $\mu$ is increased, the cost in Min model grows and in Max model is decreased.  Note that in Min model the cost can be increased up to four times as of the BTW model with ordinary drive. However, in Max model cost can be lowered down only up to two-thirds of the original model. The similar graphs for different $\alpha$'s show the same features. For $\alpha>1$ the graphs are monotonically increasing/decreasing for Min/Max model. In all cases the cost functions are convex no matter if it is Min or Max model. Increasing $\alpha$ will lead to greater changes in the cost function as we change the parameter $\mu$. Thus the BTW model with full maximum-drive has the lowest cost which  is  consistent with the previous work \cite{noel}. However implementation of maximum-drive model requires an extra cost for the local searches needed for identification of neighbors with largest heights. 

\begin{figure}
\centering

	\includegraphics[scale=.65]{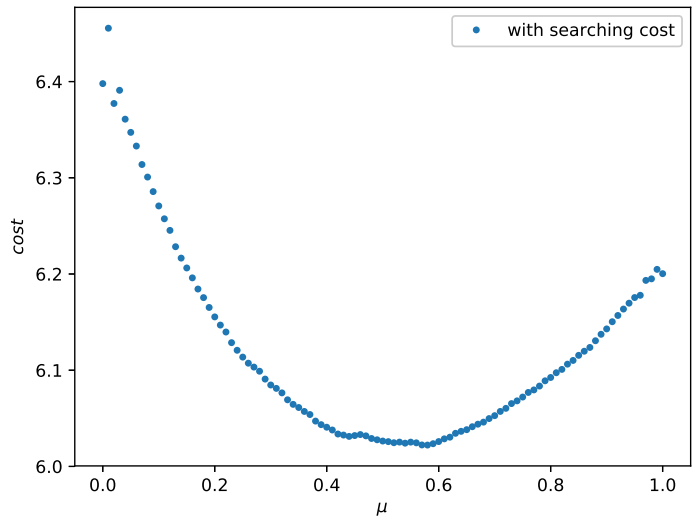}

	\label{BTWtotcost:a}

\caption{ Mean total cost of the BTW system with maximum drive implemented with probability $\mu$, Here $L=512$, $\alpha=1.5$ and $\frac{b}{c}=10^{5}$. }
\label{BTWtotcost}
\end{figure}

Let's denote the cost for each search procedure by $b$. Then it is clear that if we do the search with probability $\mu$ the cost would be $\mu b$ on average.   
Figure \ref{BTWtotcost} demonstrates total cost vs. $\mu$ for Max BTW model. In our simulations $b$ is set to 1.

 For this specific choice of parameters, the graph shows a minimum at $\mu^*\simeq 0.58$, but it is not a general feature as we will see in the following. The value of $\mu^*$, if it exists, depends on the ratio $b/c$, $\alpha$ and the size of the system $L$. 

The average cost obtained in equation (\ref{average cost}) can be interpreted in the following way. The cost is  $\langle {\rm cost}\rangle= \int c s^\alpha p_\mu(s,L) ds =  c\langle s^\alpha\rangle_\mu$ which is simply one of the moments of $s$ with respect to distribution unction $p_\mu(s,L)$. In the previous section, we have shown that with the new drive method, the system is still critical, that is, $p_\mu(s)=s^{-\tau} g_\mu(s/L^D)$ where $g_\mu$'s are the scaling functions for systems with different values of $\mu$. In large enough critical systems the moments depend on the size of the system via  $\langle s^\alpha\rangle_\mu=L^{\alpha +\tau-1} f_\alpha(\mu)$.  The function $f$ depends on the probability distribution function of the system and cannot be determined explicitly in our models, However we have already obtained it through simulations (see figure \ref{BTWcost} for the case $\alpha=1.5$). We have observed that for $\alpha>1$ the  function $f_\alpha(\mu)$ is a decreasing convex function .  On the other hand, the search cost is simply proportional to $\mu$ and is independent of size, that is, $cost_{\rm search}=b \mu$. To find the optimum value for search depth we have to solve the equation $b/c=-L^{\alpha+\tau-1} \frac{\partial f}{\partial \mu}$. This equation will have solution for $\mu$ in the interval $[0,1]$ if and only if $-f_{,\mu}(0) L^{\alpha+\tau-1} >b/c>-f_{,\mu}(1) L^{\alpha+\tau-1}$, where $f_{,\mu}= \frac{\partial f}{\partial \mu}$. It is clear that for very small sizes, the search would not be beneficial and for very large sizes one has to do full search. However, for intermediate sizes, which we have considered in figure (\ref{BTWcost})  there would be some optimum value for $\alpha$.

\begin{figure}
\centerline{\includegraphics[scale=.65]{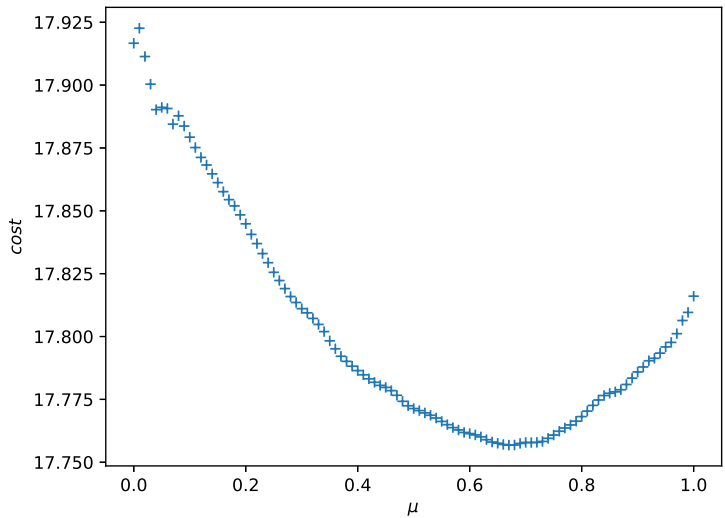}}
\caption{Mean cost as a function of $\mu$  in Max Manna model where $\mu$ is the probability that max drive is implemented. The total cost shows a minimum at $\mu\approx 0.68$. Here $L=512$ and $\alpha=1.5$ and $\frac{b}{c}=10^{5} $. }
\label{Mannatotcost}
\end{figure}

We have performed the very same steps for the Manna model and find a similar behavior. Figure \ref{Mannatotcost} shows the total cost of the Max Manna model in which the parameter $\mu$ is the probability of applying max drive. Again depending on the specific choice of parameters like $c$ and $b$ there might be a minimum total cost as it is seen in the graph. Again with the same reasoning, for large systems it is beneficial to apply the maximum drive all the times.    

\begin{figure}
\centerline{\includegraphics[scale=.45]{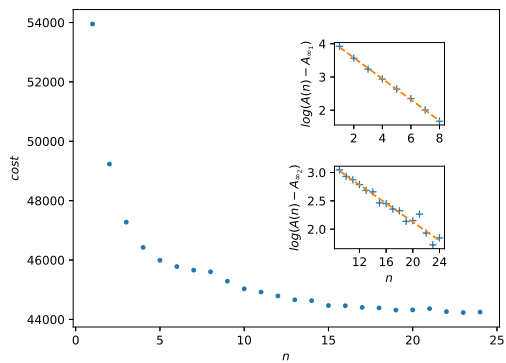}}
\caption{The avalanche cost $A(n)$ as a function of the depth of the search, $n$. Insets: $Log(A(n)-A\infty)$ shows a clear linear dependence on $n$, with different slopes for $n\leq 8$ and $8<n\leq 24$. }
\label{step}
\end{figure}

Now we turn to the question that how deep the search should be to have a better performance and least total cost. What we will do is the following: when a site is selected, $n$ of its neighboring site are randomly checked and the maximum drive is applied to these selected sites; that is, the grain of sand is added to the site with the maximum height. We want to make the changes to drive mechanism as local as possible, therefore in the case of $n\leq 8$ we only check $n$ randomly selected sites from the nearest neighbors and when $n>8$ we search the 8 nearest neighbor and a suitable number  of next nearest neighbors which are selected randomly.  The cost of avalanches  is sketched in figure (\ref{step}). In this tuning method as we increase the depth of the search, the cost of avalanches quickly saturates and deeper searches do not lower the cost significantly. It is interesting to see the behavior of the avalanche cost, $A(n)$, as a function of $n$. It seems that $A(n)$ falls exponentially to a final value $A_\infty$, that is $A(n)=A_{\infty}+(A(0)-A_\infty)\exp(-n/n_0)$. Investigating more carefully some other structure is found: for the nearest neighbors we have an exponential decay, followed by another for the next nearest neighbor. The insets of figure \ref{step} show $A(n)-A_{\infty_1}$ in semi log graphs for $n\leq 8$ (Searching only nearest neighbors) and $A(n)-A_{\infty_2}$ for $8<n\leq 24$ (Searching nearest and next nearest neighbors). Clearly the cost falls exponentially to $A_{\infty_1}=45560$ with decay constant $n_{0_1}= 3.2  $ and then with a different decay constant $n_{0_2}=12.2$ to $A_{\infty_2}=44180$. The values of the parameters are obtained using an maximum likelihood analysis. The structure continues to the third nearest neighbors but because the statistical errors are of the order of the amount of the decay, the effect is not seen as clearly as in the insets of figure \ref{step}. It worth mentioning that even if no priority is given to the nearest neighbors, the structure still can be seen but with considerable fluctuations. 

\begin{figure}
\centerline{\includegraphics[scale=.45]{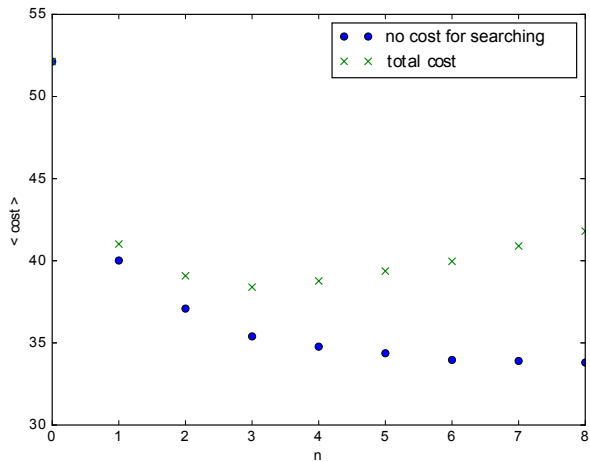}}
\caption{Total cost as a function of the number of neighbors being searched in the BTW model. The cross-curve shows a minimum at $n=3$  for $\frac{b}{c}=10^{5} $. }
\label{stepcost}
\end{figure}

Considering a search cost which is proportional to the number of sites searched, it turns out that there exists an optimum value for the search depth. For example for the parameters in figure \ref{stepcost} $n=3$ is the optimum value. It may be interesting to see how probability distribution function of avalanches varies as we change $n$. The PDF for $n=0$ and $n=9$ is essentially sketched in Fig.\ref{pdfBTW}. We observed that the PDF's for $n\gtrsim 5$ are more or less the same and like the one sketched in Fig.\ref{pdfBTW} (BTW max model).

\begin{figure}
\centerline{\includegraphics[scale=.45]{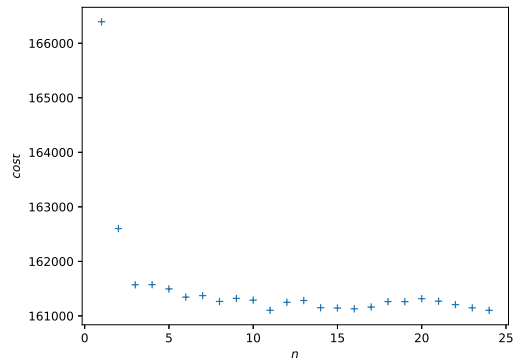}}
\caption{The avalanche cost as a function of the number of neighbors being searched in the Manna model. The cost falls down considerably for the first few points, but then the cost does not change notably. }
\label{MannaStep}
\end{figure}

We have done the same analysis with the Manna model (see figure \ref{MannaStep}). In this model the cost decreases very quickly as the depth of search is lowered, even more quickly than the case of the BTW. However, the dependence of the cost does not fit an exponential decay and the structure found in the BTW is not observed here.   

\section{conclusion}
We have considered the BTW and Manna sandpile models with locally modified drive method in which the grain of sand is added preferably to the site with Min/Max height among the neighbors. We found that in the BTW the criticality is gone with Minimum drive while the Manna model is robust to both drive methods. Also defining a cost function we have found that for large systems the new drive would be beneficial, however, there is an optimum search depth to minimize the total cost.

\end{document}